\documentclass[twocolumn,twoside,slac]{revtex4}
\usepackage{graphicx}
\usepackage{fancyhdr}
\begin{document}

\title{Calibration Infrastructure for the GLAST LAT}

\author{J. Bogart}
\affiliation{SLAC, Stanford, CA 94025, USA}

\begin{abstract}
The GLAST LAT~\cite{LATref} calibration infrastructure is designed to
 accommodate a wide range of time-varying data
types, including at a minimum hardware status bits, conversion constants,
and alignment for the GLAST LAT instrument and its prototypes. The system will
support persistent XML and ROOT data to begin with; other physical formats
will be added if necessary. In addition to the "bulk data", each data set
will have associated with it a row in a rdbms table containing metadata,
such as timestamps, data format, pointer to the location of the bulk data,
etc., which will be used to identify and locate the appropriate data set
for a particular application.

As GLAST uses the Gaudi framework for event processing, the Calibration
Infrastructure makes use of several Gaudi elements and concepts,
such as conversion services, converters and data objects and
implements the prescribed Gaudi interfaces (IDetDataSvc, IValidity, ..).
This insures that calibration data will always be valid and appropriate 
for the event
being processed. The persistent representation of a calibration dataset as
two physical pieces in different formats complicates the conversion process
somewhat: two cooperating conversion services are involved in the
conversion of any single dataset.

\end{abstract}

\maketitle

\thispagestyle{fancy}

\section{Introduction \label{intro}}
A {\bf calibration dataset} as used here is information describing some 
aspect of the state of the the GLAST LAT which may change with time.
Examples include hardware status bits, electronics gains and alignment
information. 
The {\bf Calibration Infrastructure} provides services to write and
access such datasets, particularly those needed for the earliest
stages of event processing. Information collectively known
as {\bf IRF}s (instrument response functions) used in higher-level
analyses will be managed by a different facility.

\subsection{The Instrument \label{instsrument} }
The GLAST LAT is scheduled for launch in 2006.
Although an astronomical instrument, it consists of active
components familiar to high-energy physicists: a silicon strip tracker,
a cesium iodide crystal calorimeter, and an ACD (anti-coincidence detector) 
consisting of scintillating tiles.  However, unlike the prototypical
HEP detector of today,
\begin{itemize}
\item There is no significant ambient magnetic field.
\item There is no beamline, hence no reason for circular symmetry. 
Detector components are almost all box-like and aligned 
along orthogonal axes.
\item The detetector is small, both in size and, except for silicon strips,
 in number of components.
\end{itemize}
The last especially has impact on the calibration infrastructure design. 
The amount and complexity of data to be managed is much less than in a 
typical HEP experiment, so much less that we need pay only minimal
attention to efficient use of storage and have no need for especially
clever organizational and searching schemes.

\section{Requirements \label{req}}
Although storage and cpu resources required for LAT data processing are
relatively modest, the processing
pipeline is no simpler than for standard HEP experiments; we still need
``one of everything'' as well as considerable flexibility both for developers
and for end-users.  In particular, the calibration infrastructure
\begin{itemize}
\item must support prototypes as well as flight instrument (but can assume
that no job will concern itself with more than one instrument).
\item must support remote use, even disconnected use.
\item must provide transparent access to appropriate calibration data 
for event analysis.  In particular, each calibration data set has
a {\bf validity time interval} associated with it.  If the timestamp of
the event being analyzed is outside the validity interval of the 
in-memory calibration, a new calibration whose validity interval
includes the timestamp must be fetched without special intervention
by the client.
\item must provide comprehensive, even if primitive, access for clients 
with a more diagnostic slant.
\end{itemize}
Like any other component of standard processing, the calibration infrastructure
software must be robust.  Performance and resource requirements are not
stringent, but they cannot be ignored, either.  
The calibration infrastructure should in no respect be a bottleneck.

\subsection{Software Environment \label{env} }
All offline software must build and run in at least two environments:
Linux with gcc and Windows with Visual Studio.  Developers are
split between the two.  Pipeline processing takes place
on the SLAC Linux farm.  

We prefer to use, and almost universally
do use, tools which are
both free and open source: free because we cannot ask collaborators to 
pay exorbitant per-seat charges and open source so that we can
make patches immediately ourselves if necessary.  Several of our most critical
tools have come from elsewhere within the high-energy physics community.
One such is the Gaudi framework.~\cite{GaudiRef}  The calibration 
infrastructure adheres to the Gaudi model of conversion
from persistent to in-memory (``transient'') data.

\section{Data and Metadata \label{data}}
It is extremely important that the collection of calibrations be readily
searchable by various criteria: type of calibration, validity interval,
instrument calibrated, flavor,\footnote{An application-defined string 
associated with
the calibration which may, for example, indicate intended use, such
as ``digi''. Defaults to ``vanilla''.} etc. 
However, once the desired calibration data set
is found, the chief clients of calibration
typically want access to the complete calibration data set, not just
some extracted piece.  In essence each calibration data set is comprised of two
components with different functional requirements.  One is the 
{\bf bulk data}, the calibration information about the instrument, such
as a collection of pedestals.  Since the entire collection will be
read into memory before use by most applications, the internal organization
does not have to be particularly well-matched to access patterns of the
applications, but it should be self-describing and, for large data sets,
reasonably compact.
The other component is {\bf metadata}, information {\em about} the bulk data.
The amount of metadata per calibration is small and its structure is
uniform across all calibration types so that it can be readily searched.

\subsection{Metadata \label{metadata}}
A natural storage format for the metadata is in a relational
database table.  MySQL satisfies our functional requirements and is moreover
free, open-source and straightforward to use. Each calibration is 
{\bf registered} by entering a new row in the table.  The columns in
the table fall into three categories:  those used primarily in searching
(calibration type, start and end of validity interval, and several others),
those used to access the bulk data (such as a filename and
an indicator of format), and those which contain information primarily 
intended for human readers, such as a text description of the conditions
under which the calibration was done.  See Table~\ref{table1} for the
complete list

\begin{table*}
\begin{center}
\begin{tabular}{|l|l|l|}
\hline \textbf{Field name} & \textbf{Category}& \textbf{Notes} \\
\hline serNo & search & unique index, automatically assigned \\
\hline instrument & search & identify prototype or flight instrument \\
\hline calibType & search & alignment, hot channels, etc. \\
\hline flavor & search & normally ``vanilla'' \\
\hline vstart & search & timestamp start of validity for this calibration \\
\hline vend & search & timestamp end of validity for this calibration \\
\hline completion & search & one of ok, incomplete, abort \\
\hline procLevel & search & production, development, test or superseded \\
\hline prodStart & search, info & timestamp when calib. became production \\
\hline prodEnd & search, info & timestamp when calib. was superseded \\
\hline locale    & info, possibly search & location of instrument when 
calibration was done \\
\hline enterTime & info & time at which entry was registered (automatic) \\
\hline creator & info       & procedure creating this entry \\
\hline uid     & info       & human responsible for this entry \\
\hline inputDesc & info   & comments concerning data input to calib. procedure \\
\hline notes & info & any other comments \\
\hline dataIdent & access & filename or analogous access information \\
\hline dataFmt & access & support XML and soon also ROOT \\
\hline fmtVersion & access & allows schema evolution of bulk data format \\
\hline
\end{tabular}
\caption{Metadata Table Fields} 
\label{table1}
\end{center}
\end{table*}

\subsection{Data formats \label{bulk}}
Currently only XML is used for bulk data; support for ROOT is in development.
Using a human-readable format has been an aid in debugging the new system,
but XML is not well-suited to data sets involving large numbers of 
floating-point values.  New calibrations of this nature will be written
as ROOT files as soon as full support is available.  There will be no
need to convert old files since the file format is discovered dynamically from
the metadata for each data set.

Another potential drawback of XML files compared to a binary format
like ROOT is their bulk.  If space becomes a problem support for a
third format, compressed XML, will be added.  Typical calibration files
now in use are reduced by about a factor of 100 by standard compression
techniques.

\section{Analysis Client Interface}
\subsection{Standard Gaudi Model} 
The Gaudi Detector Data Service, Persistency Service, and Conversion 
Service paradigms provide much of what is needed to meet the requirements 
of GLAST analysis clients, though only the Persistency Service could
be used untouched.  The most straightforward use of these services
might go something like this:

\begin{itemize}
\item The client asks the Data Service for calibration information while 
processing an event 
\item If data is present and current, we're done. 
\item Otherwise the Data Service requests data from the Persistency Service
\item The Persistency Service determines the appropriate Conversion Service
for data in question and makes a request
\item The Conversion Service finds the correct Converter for this
type of calibration data and invokes it
\item The Converter finds persistent form of correct data and
reconstitutes it to the appropriate in-memory location
\end {itemize}

The collaborating services pass a data structure known as an 
{\bf opaque address} among themselves which contains whatever information
is needed to identify and convert the persistent form of the dataset.
This data structure must be retrievable from the initial client request,
which typically identifies a dataset by its location in the data hierarchy
known as the {\bf TDS} (Transient Data Store).  The correct dataset is further 
implicitly determined by the timestamp of the event currently being
processed.

\subsection{Double Conversion}
The two constituents of a GLAST LAT calibration dataset are each 
converted more or less according to the above scheme.
First the metadata is converted.  
Since there is only one kind of metadata, there is no need for separate
converters; the conversion is handled directly by the MySQL Conversion
Service.  The result of this conversion is to determine the physical form
of the bulk data (e.g., XML), form a suitable opaque address incorporating
the access information from the metadata, and 
invoke the Persistency Service again with this address.  The conversion
service finds the correct converter.  In the case of XML bulk data,
a base converter class parses the XML into a DOM in-memory representation,
handles conversion of data common to all calibration types, such as
certain fields from the metadata, then invokes a method which has been
overridden by derived classes in order to handle the calibration type-specific
data.

\subsection{Transient Data Store (TDS) Organization}
\begin{figure*}
\centering
\includegraphics[width=140 mm]{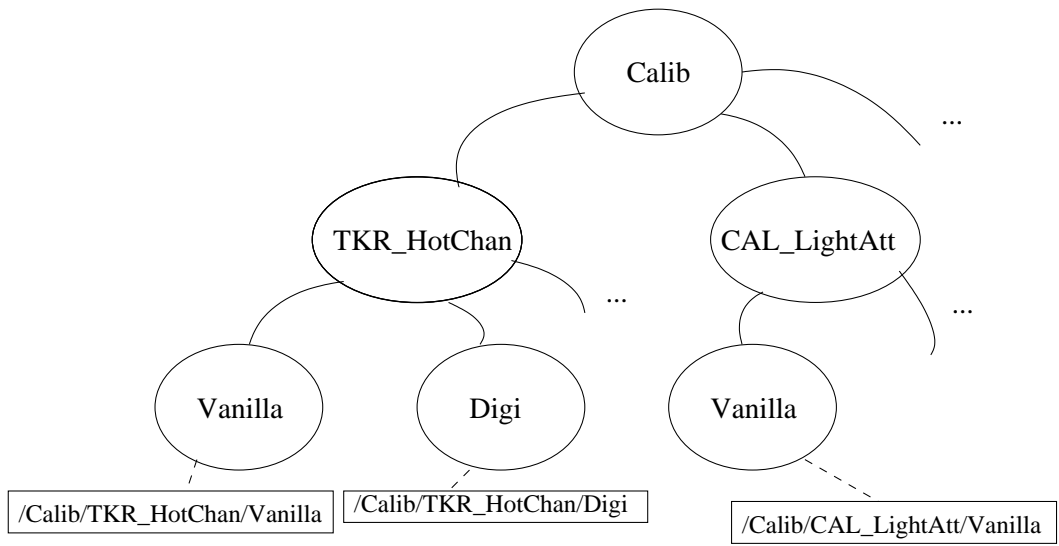}
\caption{Part of node hierarchy in TCDS. Only the bottom 
 nodes have associated calibration data.} \label{figTCDS}
\end{figure*}

Since analysis clients generally need access to an entire calibration 
dataset the TDS hierarchy (sometimes referred to as {/bf TCDS},
for Transient Calibration Data Store, to distinguish it from the
event TDS) is simple and uniform, with all the
data stored only in leaf nodes, as shown in 
Figure~\ref{figTCDS}. There is a root node, a second level
corresponding to calibration type, and a third level, the leaves,
representing flavor.  This organization allows multiple flavors of
a single calibration type to be accessible concurrently.  A ``vanilla''
leaf node is always present for each calibration type.  Other flavor
nodes are determined at run time from job options parameters.  

\section{An Example Analysis Client}
The first calibration types to be supported were dead and hot channel
lists for the tracker.  Tracker utility software already included several
Gaudi services; adding a Bad Strips Service was a natural extension.

\begin{figure*}
\centering
\includegraphics[width=140 mm]{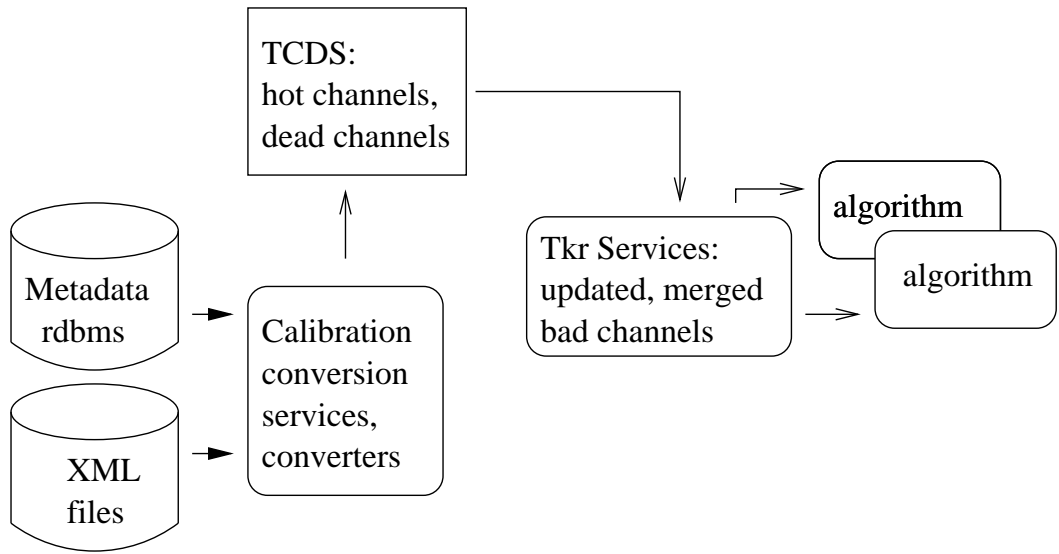}
\caption{Flow of tracker calibration data.}\label{figTkr}
\end{figure*}

Hot channels and dead channels are kept as separate datasets by the 
Calibration Infrastructure; it is not even necessarily the case that they
are generated at the same time or from the same input.  However 
analysis applications typically want a single merged list of bad
channels.  The Bad Strips Service and an associated Gaudi algorithm,
TkrCalibAlg, are responsible for the 
merging and for insuring that the list is kept up to date. 
Active elements and the flow of data in this process can be seen
in Figure~\ref{figTkr}.
For each new event,
\begin{itemize}
\item The algorithm verifies that dead and hot channel data are in 
the TCDS and causes them to be updated (to datasets whose validity
interval includes the timestamp of the current event) if necessary
\item It checks to see if the serial number (here just the serial number
entry for the corresponding MySQL row, but any unique identifier would
do) of either dataset has changed.
\item If so, it asks the Bad Strips Service to merge the new datasets.
\item Analysis algorithms may now use the Bad Strips Service to access
the merged and updated bad strips list.
\end{itemize}
An additional benefit of this design, incorporating a separate tracker
utility layer, is that the full machinery of the Calibration Infrastructure
may be bypassed
in case it is not needed, for example with MC data.

\subsection{Other clients \label{util}}
The initial focus has been on supporting analysis clients in the
Gaudi environment.
For non-Gaudi clients, tools for access to the Calibration Infrastructure 
and the data it maintains are so far rudimentary, but comprise a suitable 
base for more complete and formalized services, particularly for
access to the metadata, as the project matures.

Creation of new bulk datasets typically happens outside Gaudi and is 
not a function of the Calibration Infrastructure.
It is the responsibility of the detector subsystems to provide such data
in a supported format; that is, one which the Infrastructure conversion 
services can convert.
A new dataset becomes generally available after it has been registered
(which {\em is} an Infrastructure function) by writing a new row in the 
metadata table.
Similarly, non-Gaudi applications may use callable Infrastructure 
services to read or
search the metadata database (or may use SQL queries directly), but in 
most cases the public API for bulk data for such applications is 
just the format of the data.

\section{Improvements and Extensions}
\subsection{Metadata management}
For a robust production environment, additional tools to access and
manage the metadata will be necessary. Write access to the database is
already restricted, but even legitimate writers can make mistakes. 
Several kinds of errors can and should be caught.  Consistency 
and completeness checks
can catch more subtle problems involving more than one row.

\subsection{Production database alternatives}
The production MySQL database and bulk data at SLAC are not 
conveniently accessible for all users.  We plan to provide mirrors
for production European users.  Since the dataIdent field 
of the metadata (which
identifies the bulk data belonging to a particular calibration)  may 
and typically does contain environment variables, the metadata itself
may be left untouched in the mirrors.

Isolated developers present a somewhat different problem.  They may be
disconnected from the network entirely, but their needs are typically
less: they don't
need access to all calibrations.  One way, already successfully
demonstrated, to provide a fully-local
calibration infrastructure is to install a MySQL server on the user's
machine, fill it with a dump of the contents of the production database,
and copy over bulk data files as needed.  Another possibility which
avoids the complex process of MySQL installation would be to
provide an alternate implementation of the metadata storage and access, 
such as a simple ascii file representation of the metadata for a minimal 
collection of calibrations.

\subsection{Additional calibration types and formats}
So far only a fraction of the total number of expected calibration types
are supported by the Calibration Infrastructure; others will be
implemented as required by clients.

ROOT is the preferred choice for calibration 
files involving many floating point numbers, both because it is a
compact binary format and because it is already heavily used
for persistent event data and as an analysis platform.  
A ROOT Conversion Service is under development.

\subsection{Parameters database}

GLAST offline has as yet no formal mechanism for keeping track of 
the many parameters that go into a particular software run.  
Blessed collections of such parameters share many attributes with
calibrations.  The Calibration Infrastructure
or a near-clone could be used for this purpose.

\vspace{4mm}
\section{Conclusions}

The Calibration Infrastructure is in active use for several tracker
and calorimeter calibrations.  There is much to do before the
full facility as envisioned is in place, and certain features of even
the current limited implementation have not been thoroughtly exercised,
but there is no reason
to doubt that the design will accommodate all 
currently-anticipated future needs.

\begin{acknowledgments}
The author wishes to thank Andrea Valassi (CERN) for guidance in
the design of the two-stage conversion process and
Leon Rochester (SLAC) for suggesting the concept of \textbf{flavor} and
for contributing to the design and implementation of the bad strips
architecture.

Work supported by Department of Energy contract DE-AC03-76SF00515.
\end{acknowledgments}



\end{document}